\documentclass[12pt,reqno]{amsart}
\allowdisplaybreaks[4]
\usepackage{graphicx} 
\usepackage{amssymb}
\usepackage{amsmath}
\usepackage{cite}

\begin{document}
\title[Integrability ]{Designable integrability of the variable coefficient nonlinear Schr\"odinger
equation}
\author{Jingsong He* \dag, Yishen Li\ddag
} \dedicatory { \dag \ Department of Mathematics, Ningbo University,
Ningbo , Zhejiang 315211, P.\ R.\ China \\
 \ddag \ Department of Mathematics, USTC, Hefei,Anhui 230026  ,
P.\ R.\ China }
\thanks{$^*$ Corresponding author: hejingsong@nbu.edu.cn,jshe@ustc.edu.cn}
\begin{abstract}

The designable integrability(DI)\cite{definition} of the variable coefficient nonlinear Schr\"odinger equation
(VCNLSE) is first introduced by construction of an explicit transformation which maps VCNLSE to the usual
nonlinear  Schr\"odinger equation(NLSE). One novel feature of VCNLSE with DI is that its coefficients can be
designed artificially and analytically by using transformation. A special example between nonautonomous NLSE
and NLSE is given here. Further, the optical super-lattice potentials (or periodic potentials)
and multi-well potentials are designed, which are two kinds of important potential in Bose-Einstein
condensation(BEC) and nonlinear optical systems. There are two interesting features of the soliton of the
VCNLSE indicated by the analytic and exact formula. Specifically, its the profile is variable and
its trajectory is not a straight line when it evolves with time $t$.
\end{abstract}

 \maketitle


\noindent
\section{Introduction}
The variable coefficient soliton equations with different
formulations have been well studied since 1970s, and there are many
results on this topic in the literatures. For example, inverse
scattering method \cite{cl,calgoero1, newell},symmetry
algebra\cite{li1}, etc. However, the solving and applying of the
several types of the variable coefficient nonlinear Schr\"odinger
equation(VCNLSE) has been revived recently in the research of
Bose-Einstein condensation(BEC) and nonlinear optics
\cite{akhmediev,kivshar1,vnsah1,vnsah2,kpd,agrawal,kp1,jvv1,jvv2,su,liu1,sivan}.
It is well known that the main difficulty in solving process is due
to the non-isospectrum property in the Lax pair of the integrable
VCNLSE.  So it is very natural to get the idea that one can overcome
this difficulty by finding a transformation mapping the VCNLSE to
the usual nonlinear Schr\"odinger equation(NLSE) associated with an
iso-spectral problem, and thus one can get plenty of solutions of
the VCNLSE from the abundance of known solutions of the NLSE.  This
idea is pointed out at the added note of the  very early
work\cite{cl} on the NLSE with a linear potential of $x$, which is
solved by inverse scattering method. Since then, in order to get the
explicit solutions of  some special cases of VCNLSE with concrete
coefficients, there are several typical transformations involving
dependent variables and independent variables, such as lens type
transformations\cite{ss,kontop1,zhangjiefang1,zhangjiefang2},
similarity transformations
\cite{vnsah1,vnsah2,kpd,agrawal,jvv2,konotop2,jbb1,chenlin} and some
others\cite{he1,ysli,Gurses,luo1,luo2, kundu}. However, for the
extensively studied potentials in BEC and nonlinear optical
systems,i.e. optical-lattice and super-lattice potentials (or
periodic
potentials)\cite{kivshar2,malomed1,konotop3,kevrekidis1,wubiao} and
multi-well
potentials\cite{kevrekidis1,kivshar3,malomed2,ketterle1,carr,kevrekidis2},
to our best knowledge, the exact and analytical solutions of the
VCNLSE such as solitons, have not been reported in the literatures
except stationary solutions and numerical solutions. In order to
show dynamical properties of the solitons in one-dimensional BEC and
nonlinear optical systems under the control of the two kinds of the
external potentials above mentioned, we shall develop a new method
to design corresponding integrable VCNLSE, and then get its soliton
solutions.

  The organization of this paper is as follows.  In section 2, a general method which maps VCNLSE to NLSE is
given with several arbitrary functions. These arbitrary functions provide a possibility  to design integrable
model. In section 3, as a special application of this method, the nonautonomous NLSE\cite{vnsah2} is designed
and mapped explicitly to NLSE. In section 4, two NEW integrable models with important physical concerns
are designed, and their properties are discussed according to analytic solutions given from a single soliton of
the NLSE by means of our general method. The conclusion will be given in section 5.

\section{General Method}

The investigation object of this article is a very universal  VCNLSE
in the form of
\begin{equation}\label{eqvcnlse}
i\dfrac{\partial }{\partial
t}\psi(x,t)+\frac{1}{2}o\dfrac{\partial^2}{\partial x^2}\psi(x,t)-
v\psi(x,t)-g|\psi(x,t)|^2\psi(x,t)=0,
\end{equation}
where $o=o(x,t), v=v(x,t), g=g(x,t)$ are three real functions of $x$
and $t$. This equation is a light extension of eq.(1) in reference
\cite{jvv2} when $o\not =1$, which is widely used to characterize
one-dimensional BEC and nonlinear optical systems under some
physical conditions. First focus on the integrability of
eq.\eqref{eqvcnlse} by means of looking for a direct relationship
between VCNLSE eq.(\ref{eqvcnlse}) and usual    nonlinear
Schr\"odinger  equation (NLSE)  eq.\eqref{standardnls}.
Then, find  a  transformation  mapping eq.\eqref{eqvcnlse} to the usual NLSE,
\begin{equation}\label{standardnls}
i\frac{\partial q}{\partial T}+\frac{1}{2}\frac{\partial^2
q}{\partial X^2}+e |q|^2q=0,
      \end{equation}
where $e=\pm 1$, $q=q(X,T)$. Meanwhile, coefficients $o,v,g$ are
given analytically by this transformation. Therefore, one advantage
of this transformation is to solve VCNLSE by using all known
solutions of NLSE as we discussed  in the above section.

 To this purpose, a trial transformation
\begin{equation}\label{mapforvcnlse}
\psi(x,t)=q(X,T)p(x,t)e^{i\phi(x,t)}
\end{equation}
is introduced with $X=X(x,t),T=T(t)$. So the central task is to determine the
concrete expressions of real smooth functions \{$o,v, g, X, p, \phi$
\} by requesting $q(X,T)$ to satisfy  the standard NLSE
eq.(\ref{standardnls}). By substituting the transformation
eq.(\ref{mapforvcnlse}) into eq.(\ref{eqvcnlse}),
 and setting
\begin{equation}\label{eqog1}
o(x,t)=\dfrac{ T_t} {( X_{x} )^2 }, g(x,t)= -\dfrac{e T_t }{p^2}
\end{equation}
without loss of generality of the transformation, then  it becomes
\begin{align}\label{transformedeqbyki}
&  iq_T + \dfrac{1}{2}q_{XX}+ e |q|^2q+\dfrac{1}{2} \dfrac{(i k_3
+k_4)q}
 { p T_t ( X_x)^2}
 +\dfrac{1}{2} \dfrac{(i k_1 +k_2) q_X}{ p T_t(  X_x)^2}=0.
\end{align}
Note $q_X$ denotes $\dfrac{\partial q}{\partial X}$, $T_t$ denotes
$\dfrac{d T}{d t}$ and so on. Here $k_i$, $i=1, 2, 3,4$, are given
by
\begin{align}
&k_1= 2p X_t X_x^2+2p T_t\phi_x X_x,\ \  k_2= T_t  \big(2p_xX_x+ p
X_{xx} \big), \nonumber \\
 &k_3=2p_t X_x^2 +T_t\big( 2p_x \phi_x+p
\phi_{xx}\big),\ k_4=T_t p_{xx} -2p \phi_t X_x^2 -T_t p \phi_x^2- 2v
p X_x^2.        \nonumber
\end{align}
 Obviously,
let $ k_1=k_2=k_3=k_4=0$, then eq.(\ref{transformedeqbyki}) becomes
the standard NLSE. Moreover, we get
\begin{align}
&X=X(x,t)= \int F(x)f_1(t)dx+f_3(t),  \quad T=T(t), \label{eq1parametersforvcnlse}\\
&p=p(x,t)=\dfrac{f_1(t)}{\sqrt{F(x)f_1(t)}},  \label{eq2parametersforvcnlse} \\
&\phi=\phi(x,t)=-\int \dfrac{\big( \int F(x)f_{1t} dx +f_{3t}
\big)F(x)f_1(t) }{T_t} dx+ f_2(t),
  \label{eq3parametersforvcnlse}\\
&v(x,t)= -\dfrac{1}{8}\dfrac{-3T_t  F_x^2+2T_tF_{xx}F +8f_1(t)^2
\phi_tF^4+4T_t(\phi_{x})^2F^2  }{f_1(t)^2F(x)^4} ,
 \label{vparameters}
\end{align}
from $ k_1=k_2=k_3=k_4=0$ by a  tedious calculation. Furthermore,
taking  $X,p,T$ back into eq.(\ref{eqog1}), then
\begin{equation}
o(x,t)=\dfrac{T_t}{ F(x)^2f_1(t)^2}, \quad
g(x,t)=-\dfrac{eF(x)T_t}{f_1(t)}. \label{oandgparameters}
\end{equation}
So the transformation  in  eq.(\ref{mapforvcnlse}) indeed maps the
VCNLSE to the NLSE as we wanted, which is determined by
eq.(\ref{eq1parametersforvcnlse}) to eq.(\ref{oandgparameters})
through five real arbitrary
 functions  $f_1(t),f_2(t), f_3(t), T(t), F(x)$,  $F(x)f_1(t)>0$.
 At last, we would like to point out that  we may set $X=X(x,t)$ and  $T=T(x,t)$ in transformation
 eq.(\ref{mapforvcnlse}) for better universality. However, to eliminate the term $(\dfrac{\partial^2}{\partial
 X\partial T} q(X,T))$ in the transformed equation, we have to ask $(\dfrac{\partial }{\partial x} X )=0$ or
$(\dfrac{\partial }{\partial x} T )=0$. So we choose directly
$T=T(t)$ for simplicity in eq.(\ref{mapforvcnlse}). The other choice
$X=X(t),T=T(x,t)$ will be given in a separate paper.

Thus we can design $o$, the external potential $v$ and interaction
nonlinearity $g$  according to different physical considerations by
means of the selections of the arbitrary functions above mentioned,
such that the integrability is guaranteed. Therefore, we call that
the VCNLSE possesses the designable
integrability(DI)\cite{definition}, which originates from the rigid
integrability\cite{definition} of the NLSE and  the transformation
eq.(\ref{mapforvcnlse}).
\section{Nonautonomous Nonlinear Schrodinger Equation}

By comparing with the known results on the connection between VCNLSE
and NLSE  mentioned at the first paragraph, our result is  more
universal because the research equation eq.(\ref{eqvcnlse}) and
transformation eq.(\ref{mapforvcnlse}) is more general.  Moreover,
the integrable conditions \cite{vnsah1,vnsah2,ysli,Gurses,luo1,luo2}
of the coefficients in the VCNLSE disappear in our method. Of
course, to guarantee the integrability of VCNLSE, these coefficients
can not be arbitrary functions. Actually, these conditions are
satisfied automatically by analytical expressions of coefficients
$o,\ v$ and $g$.

To show this point clearly, we would like to demonstrate that the
nonautonomous NLS\cite{vnsah2} has designable integrability. In
other words, we shall show how to choose functions
$f_1(t),f_2(t),f_3(t)$, $T(t),F(x)$ such that we can get
nonautonomous NLSE \cite{vnsah2}
\begin{eqnarray}\label{nonatuognls}
i\frac{\partial Q}{\partial t}+\frac{ D(t)}{2}\frac{\partial^2
Q}{\partial x^2} +e R(t)|Q|^2Q-2\alpha(t)x
Q-\frac{\Omega^2(t)}{2}x^2Q=0, \quad e=\pm 1,
\end{eqnarray}
 with a integrability condition
\begin{eqnarray}\label{nonatuognlscondition}
-\Omega^2(t)D(t)=\frac{d^2}{dt}\ln
D(t)+R(t)\frac{d^2}{dt^2}\frac{1}{R(t)}
  -\frac{d}{dt}\ln D(t)\frac{d}{dt}\ln R(t),
\end{eqnarray}
from VCNLSE, and nonautonomous NLSE can be transformed to the usual NLSE
by previous transformation
 eq.(\ref{mapforvcnlse}). So by comparing with nonautonomous NLSE,
then $g(x,t)=-e R(t),o(x,t)=D(t), v(x,t)=
v_2x^2+v_1x+v_0,v_2=\dfrac{\Omega^2(t)}{2}, v_1=2\alpha(t), v_0=0$.
To get this integrable model, setting  $F(x)=F_0$, and  using
$o(x,t)$ and $g(x,t)$ we get
\begin{equation}\label{c1teq}
f_1(t)=\dfrac{ R(t)}{ D(t) F_0^3},
\end{equation}
\begin{equation}\label{Teq}
T(t)=\int\dfrac{ R(t)^2}{D(t) F_0^4 }    dt+ T_0.
\end{equation}
Here $F_0$ is a real constant, $T_0$ is a integral constant. Taking
$F(x)=F_0$ into the expression of $v(x,t)$ in
 eq.(\ref{vparameters}),
it infers that $v(x,t)$ is a second order polynomial of $x$. Then
\begin{equation} \label{c2teq}
f_2(t)=-\dfrac{1}{2} \int \dfrac{ \Big(2\int \dfrac{ D(t) \alpha(t)
} { R(t)}dt- c_{301}  \Big)^2  R(t)^2} {D(t)  }    dt+c_{20}
\end{equation}
and
\begin{equation}\label{c3teq}
f_3(t)=\int \dfrac{ \Big( 2\int \dfrac{ D(t) \alpha(t) }{ R(t)}
  dt- c_{301}  \Big)  R(t)^2}
{ F_0^2D(t) }    dt+c_{302}
\end{equation}
are given from eq.(\ref{vparameters}) with the help of
$v_1=2\alpha(t), v_0=0$. Here $c_{20}, c_{301},c_{302}$ are integral
constants. Moreover,
 taking $v_2=\frac{1}{2} \Omega(t)^2$ in  $v(x,t)$ infers
\begin{eqnarray}\label{integrablecondition}
\dfrac{1}{2}\dfrac{R_t D_t  }{R D^2} -\dfrac{R_t^2 }{R^2 D}
-\dfrac{1}{2}\dfrac{ D_{tt}  } {D^2} +\dfrac{1}{2}\dfrac{ R_{tt}  }
{R D} +\dfrac{1}{2} \dfrac{D_t^2 }{D^3}=\dfrac{1}{2}\Omega^2,
\end{eqnarray}
which is equivalent to  eq.(\ref{nonatuognlscondition}). Further,
taking $F(x)=F_0$ and $f_1(t)$ given by eq.(\ref{c1teq}) into
eq.(\ref{eq2parametersforvcnlse}), then
\begin{equation}\label{peq}
p=p(t)=\dfrac{1}{F_0^2}\sqrt{\dfrac{R(t)}{D(t)}}.
\end{equation}
According to eq.(\ref{eq3parametersforvcnlse}) and
eq.(\ref{eq1parametersforvcnlse}), and using known functions
$f_1(t),f_2(t),f_3(t),T(t),F(x)$, then
\begin{eqnarray}
& X=&F_0 f_1(t)x+f_3(t), \label{Xeq}\\
& \phi=&b_2(t)x^2+b_1(t)x+b_0(t),  \label{phieq}\\
& b_2=&-\dfrac{1}{2}\dfrac{( R_t D-R D_t)}{R D^2}   \\
& b_1=&\dfrac{R(t)\left(-2\int\dfrac{D(t)\alpha(t)}{R(t)}+c_{301}  \right)  }{D(t)}  \\
& b_0=&f_2(t)=-\dfrac{1}{2} \int \dfrac{ \Big(2\int \dfrac{ D(t)
\alpha(t) } { R(t)}dt- c_{301}  \Big)^2  R(t)^2} {D(t)  } dt+c_{20}.
\end{eqnarray}
We have verified that the transformation
\begin{equation} \label{mapautonomousnlstonls}
Q(x,t)=q(X,T)p(t)exp(i\phi)
\end{equation}
given by eq.(\ref{Teq}),eq.(\ref{peq}),eq.(\ref{Xeq}) and
eq.(\ref{phieq}) indeed maps nonautonomous NLSE
eq.(\ref{nonatuognls}) to the usual NLSE. It is trivial to find that
transformation eq.(\ref{mapautonomousnlstonls}) is equivalent to the
result in Ref. \cite{Gurses}.

    As the end of this section, we would like to show several examples of nonautonomous NLSE eq.(\ref{nonatuognls}),
and integrable condition Eq.(\ref{nonatuognlscondition}) held for
them. Their solutions can be obtained from known solutions of the
NLSE by transformation Eq.(\ref{mapautonomousnlstonls}). In
particular, all of the following equations has DI property.

\noindent Example 1 \cite{liu1}:\ \ \  Let $D(t)=2, R(t)=2 g_0
e^{\lambda t},\Omega(t)^2=-\frac{\lambda^2}{2}, \alpha(t)=0$, then
the nonautonomous NLSE reduces to Eq.(1) of reference \cite{liu1} with $a(t)=g_0e^{\lambda
t}$.

\noindent Example 2 \cite{liu2} : \ \ \ Let $D(t)=1,
R(t)=1,\Omega(t)^2=0, \alpha(t)=\frac{1}{2}(B+C \sin (wt))$, then
the nonautonomous NLSE reduces to Eq.(2) of reference \cite{liu2}.

\noindent Example 3 \cite{xue1}:\ \ \ Let $D(t)=2, R(t)=1+\tanh
(wt), \Omega(t)^2=(\tanh (wt) -1)w^2, \alpha(t)=0$, then the
nonautonomous NLSE reduces to Eq.(1) of reference \cite{xue1} with $g(t)=1+\tanh (wt)$ and $
k(t)= 1-\tanh(wt)$.

\noindent Example 4 \cite{su,zhangjiefang1}: \ \ \ Let $D(t)=1,
R(t)= 1+m \sin(wt),\Omega(t)^2=2k(t), \alpha(t)=0$,then the
nonautonomous NLSE reduces to Eq.(2) of
reference\cite{zhangjiefang1} with $g(t)=1+m sin(w t)$ and
$k(t)=-\dfrac{1}{2}\dfrac{mw^2( 2m\cos(wt)^2 +sin(wt) +m
\sin(wt)^2)}{1+2m \sin(wt)+m^2\sin(wt)^2}$.

\noindent Example 5 \cite{zhangjiefang3}: \ \ \ Let $D(t)=1, R(t)=
1,\Omega(t)^2=0, \alpha(t)=-f(t)/2$ ,then the nonautonomous NLSE
reduces to Eq.(3) of reference\cite{zhangjiefang3} with $\eta=-1$
and $f(t)=b_1+l\cos \omega t$ given by eq.(30) of
reference\cite{zhangjiefang3}.

\section{New Examples}

To further illustrate the  wide applicability  of our methodology,
and motivated by the extreme importance of the external potentials
in  the BEC and nonlinear optics systems,
two NEW integrable models:  integrable VCNLSE with optical super-lattice potentials
(or periodic potentials) and multi-well potentials are designed respectively. In
the following examples, we shall set $f_1(t)=1,f_2(t)=1,f_3(t)=1,
T(t)=t,e=1$.\\
{\noindent\bf New Example 1: Optical super-lattice potentials}
According to transformation eq.(\ref{mapforvcnlse}), we have three
potentials:1)$ F(x)=1/(\cos (x)+3/2 ), v(x,t)= -\dfrac{3}{16}-
\dfrac{1}{16}\cos (2 x)-\dfrac{3}{8} \cos (x)$; 2)$ F(x)=1/(\cos
(2x)+\cos (x)+21/10 ), v(x,t)= -2 \cos(x)^4-\dfrac{3}{2}\cos
(x)^3-\dfrac{93}{40}\cos (x)^2
 -\dfrac{11}{40} \cos (x) + \dfrac{39}{40}$;
 3)$ F(x)=3/(\cos (3x)+\cos (2x)+5 \cos (x)+10 ), v(x,t)=-2\cos (x)^6-\dfrac{4}{9}\cos (x)^5 -
\dfrac{10}{9}\cos (x)^4- \dfrac{80}{9}\cos (x)^3-\dfrac{25}{18}\cos
(x)^2+\dfrac{11}{2} \cos (x) + \dfrac{17}{18}$ from
eq.(\ref{vparameters}). They are real periodic  functions of $x$
with period $2\pi$,
 and have one peak, two or three peaks over intervals of length $2\pi$ respectively. The profiles of them are  plotted in Figure 1 from left to
right in order.
\begin{figure}[htbp]
      \centering \mbox{}\hspace{-1cm}
      \includegraphics[width=0.5\textwidth]{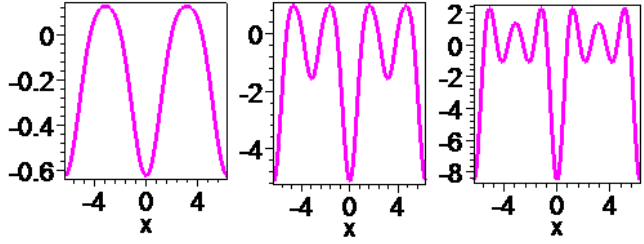}
\caption{{
 The designed optical super-lattice potentials({\bf New
Example 1})
 with period $2\pi$.} }\label{periodicpotential}
\end{figure}

\noindent{\noindent\bf New  Example 2: Multi-well potentials}
According to transformation eq.(\ref{mapforvcnlse}), we have got
three types of multi-well potentials: 1) Let $a>0$, then
$F(x)=\dfrac{1}{200}\dfrac{1}{\cos({\rm arccot}((x/20)^2+a))}$ gives
symmetric double well potentials, which are  plotted in Figure
2(left) for $a=0.001,0.2$ and $1$; 2)Let $a>0$, then
$F(x)=((\dfrac{x}{30})^4+a (\dfrac{x}{50})^2+ (\dfrac{x}{80})+4)/400
$ gives non-symmetric double well potentials, which are  plotted in
Figure 2(middle) for $a=1/15,2$ and $7$. Note that the last one is a
single well potential. 3) $F(x)=\dfrac{1}{15}\dfrac{1}{2\cos(4 {\rm
sech}(\frac{x}{20}))+ 1.3\cos({\rm sech}(\frac{x}{20})) +10} $ gives
triple well potential, which is plotted in Figure 2(right).
\begin{figure}[htbp]     \centering \mbox{}\hspace{0cm}
\includegraphics[width=0.5\textwidth]{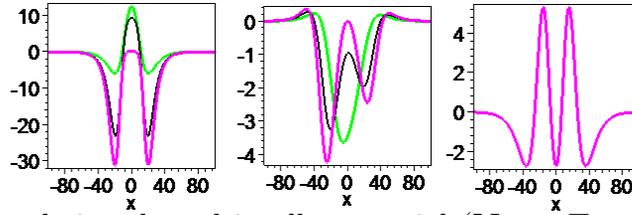}
 \mbox{}\vspace{-0.5cm}\caption{{
  The designed multi-well  potentials({\bf New Example 2}). Left panel(the
  symmetric double well potentials): From the thinnest(magenta) to the thickest(green)
  curves it denotes potential generated by $F(x)$ with $a=0.001,0.2,1$,
  respectively. Middle panel(the non-symmetric double well
  potentials): From the thinnest(magenta) to the thickest(green) curves it denotes
  potential generated by $F(x)$ with $a=1/15,2,7$,respectively. Right
  panel:the triple well potential.}} \label{figp1sol1-2d3d}
 \end{figure}

\noindent Note that the associated other two coefficients $o(x,t)$
and $g(x,t)$ are given simultaneously by means of transformation
eq.(\ref{mapforvcnlse}), such that VCNLSEs associated with these
designed coefficients $o,g$ and $v$ are integrable systems.
To save space, we do not write out them here. Moreover, double well potential can be achieved
by taking $F(x)=0.35
x^2+5{\rm sech}^2(0.27x)$.  Additionally, by setting arbitrary functions in transformation
eq.(\ref{mapforvcnlse}), the interested readers can design different
integrable VCNLSE to establish mathematical model equation of
physical systems. For instance, time-dependent  optical
super-lattice potential\cite{kivshar4} can also be designed by
setting $T=\int f(t)dt$ and $F(x)=1/(a\cos (x)+ b)$ in $v(x,t)$.
Here $b>a>0$.

Furthermore, the transformation eq.(\ref{mapforvcnlse}) provides an efficient way to construct exact and analytic
solutions $\psi(x,t)$ of the designable VCNLSE from known solutions $q(X,T)$ of the NLSE, such that we can explore
the dynamical evolution of solitons of the VCNLSE conveniently. For example, setting a usual single soliton  $q(X,T)$  as
\begin{equation}\label{singlesolitonNLSE}
|q(X,T)|=\frac{2\sqrt{2}|\eta_1|}{\cosh (2\eta_1 \sqrt{2}X+8\eta_1\xi_1T)},
\end{equation}
which is given by eq.(6) of reference \cite{he1}.
Obviously, the profile of the usual soliton $q(X,T)$ on the plane of (X,T) is invariant when soliton evolves
with time $t$. A single soliton of the VCNLSE,
\begin{equation}\label{singlesolitonVCNLSE}
|\psi(x,t)|=|p(x,t)||q(X,T)|=|p(x,t)|\frac{2\sqrt{2}|\eta_1|}{\cosh (2\eta_1 \sqrt{2}X+8\eta_1\xi_1T)},
\end{equation}
is obtained by using transformation eq.(\ref{mapforvcnlse}).
Note that the profile of the soliton $|\psi(x,t)|$ of the VCNLSE is not preserved when it evolves with time
$t$ because the amplitude $|p(x,t)|$ is a function of $x$ and $t$  and the trajectory is a curve $x=x(t)$ on
the plane of (x,t), which is defined implicitly by
\begin{equation}\label{trajectoryvcnlse}
\sqrt{2}X(x,t)+4\xi_1T(t)=0.
\end{equation}
This shows that the profile of the soliton of the VCNLSE is designable by using different $p(x,t)$, $X(x,t)$ and
$T(t)$, which can be realized by choosing different arbitrary functions $f_1(t),f_3(t),F(x)$ and $T(t)$ in
transformation eq.(\ref{mapforvcnlse}). In particular, as we shall show in the following example,
the amplitude of $\psi(x,t)$ is dependent of $t$ even if the  $p=p(x)$ is x-dependent only, because $p=p(x(t))$ when soliton
moves along the trajectory $x=x(t)$.

To further illustrate the property of dynamical evolution of the soliton $|\psi(x,t)|$ of the VCNLSE,
one example is given here. According to the above formula eq.(\ref{singlesolitonVCNLSE}),
the solution of the case 1) in {\bf New Example 1} is a deformed  single soliton,i.e.,
\begin{equation}\label{solitonvcnls}
|\psi(x,t)|=\dfrac{2|\eta1|}{\sqrt{\dfrac{1}{2\cos(x)+3}} \cosh\Big(2\sqrt{2}\eta_1 \widetilde{X}
+8\xi_1\eta_1 t \Big)},
\end{equation}
which is plotted in Figure 3  with $\xi_1=0.1$ and $\eta_1=-0.1$.
\begin{figure}[htbp]     \centering \mbox{}\hspace{0cm}
      \includegraphics[width=0.25\textwidth]{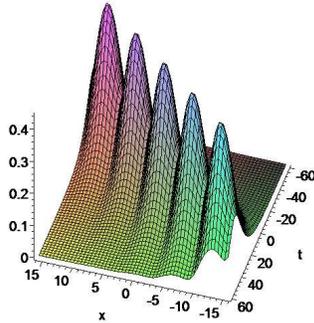}
      \mbox{}\vspace{-0.5cm}\caption{{\footnotesize
      The dynamical evolution on ($x-t$) plane of a single soliton of the
      VCNLSE with an optical super-lattice potential designed by case 1) of {\bf New Example1}.} }
\end{figure}
Here  $\widetilde{X}= \dfrac{4}{\sqrt{5}}\big(\arctan(\dfrac{1}{\sqrt{5}}\tan (\dfrac{x}{2}) )+
 \big[\dfrac{\frac{x}{2} +\frac{\pi}{2} }{\pi}\big]\pi\big) +1$, and $\big[x\big]$ is the greatest integer less than or equal
 to $x$. Particularly, $\widetilde{X}$ is a monotonically increasing and continuous function of $x$, which is
 obtained from eq.(\ref{eq1parametersforvcnlse}) by adding the floor function $\big[x\big]$. Figure 3 shows
intuitively two interesting features of the soliton of the VCNLSE: 1) the profile is variable and 2) the
trajectory is not a straight line  when it evolves with time $t$, as we pointed in the above paragraph. This is also
supported visibly by Fig.4 for profiles of different time $t$ and Fig 5 for the trajectory on the plane of (x,t),
which is given by
$
\sqrt{2} \widetilde{X}+ 0.4 t=0.
$

\begin{figure}[htbp]     \centering \mbox{}\hspace{0cm}
      \includegraphics[width=0.25\textwidth]{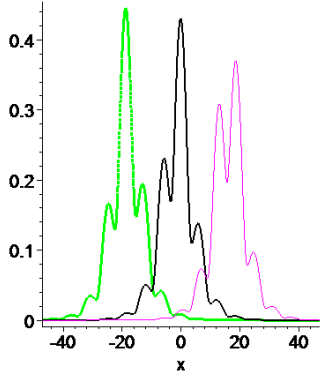}
      \mbox{}\vspace{-0.5cm}\caption{{\footnotesize
       The one-dimensional profiles of single soliton in Fig.3 at different
      times. From the right(magenta) to the left(green) curves it denotes soliton at  $t=-55,0,55 $,
      respectively.     }
      }
\end{figure}

\begin{figure}[htbp]     \centering \mbox{}\hspace{0cm}
      \includegraphics[width=0.25\textwidth]{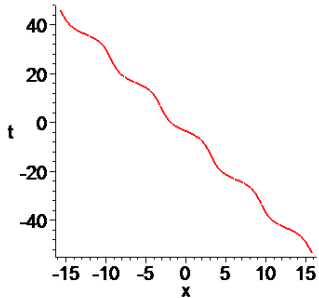}
      \mbox{}\vspace{-0.5cm}\caption{{\footnotesize
Trajectory of the single soliton in Fig 3.
      }
      }
\end{figure}

Note that there are many other kinds of solutions of NLSE, such as dark
soliton, periodic solution, position, negation, complexiton, etc.,
which can be used to generate the solutions of the VCNLSE.
In ourexamples, merely the single bright soliton of the NLSE is used.
Especially, from the ``seed"  of the multi-soliton solutions of
NLSE, the multi-soliton solutions of the VCNLSE can be obtained,
 and then its interaction properties are discussed. However, it is very difficult to
find multi-soliton solutions of the VCNLSE from the view of
non-isospectral integrable system.

\section{Conclusions}
In conclusion, a new concept ``designable integrability "\cite{definition} of the VCNLSE has been developed
by a novel transformation eq.(\ref{mapforvcnlse}). The novel characteristic of VCNLSE with DI  is that the
variable coefficients with physical meaning including  $o(x,t),v(x,t)$ and $g(x,t)$ can be designed
artificially  and analytically according to different physical considerations. Correspondingly, the profile of
its solutions including soliton and others are also tunable intentionally by using different $p(x,t)$, $X(x,t)$
and $T(t)$, which can be realized by choosing different arbitrary functions.  The
nonautonomous NLSE\cite{vnsah2} is re-obtained as a special design
through choosing $F(x)=F_0$(constant). Furthermore, two kinds of NEW VCNLSE with optical super-lattice potentials
and multi-well potentials have been designed respectively, and two interesting features of the soliton
of the VCNLSE are shown by formulas and figures. The results in this paper show that DI and some unusual
behaviors of solitons for the VCNLSE originate from the usual NLSE and the transformation
eq.(\ref{mapforvcnlse}). The methodology of studying  variable
coefficients partial differential equations with DI can be extended
to many other cases including other (1+1)-dimensional, even
higher dimensional integrable systems, multi-component systems
and discrete systems. In particular, the method to design the
solvable $o$, $v$ and $g$ of the VCNLSE is expected optimistically
to be used by theoretical and experimental researchers.

As the end of this paper, the method applied to  a more general VCNLSE
\begin{equation}\label{eqgvcnlse}
i\dfrac{\partial }{\partial
t}\psi(x,t)+\frac{1}{2}o\dfrac{\partial^2}{\partial x^2}\psi(x,t)-
v\psi(x,t)-g|\psi(x,t)|^2\psi(x,t)-i\Gamma \psi(x,t)=0,
\end{equation}
which is a minor extension of modeling  equations of BECs and
nonlinear optics in recent many works,is shown here. In order to guarantee
integrability of eq.(\ref{eqgvcnlse}), Luo et al have shown that
$o,g, \Gamma$ must be time-independent, and $v(x,t)$ must be a
second order polynomial of $x$ and satisfy a integrable
 condition (eq.(22) of reference\cite{luo2}). However, their integrable conditions  are too strong
 because of the restriction of the Painleve analysis, and a very general
forms of $o,g,v$ , which are real functions of $x$ and $t$, has been achieved by a
similar transformation of eq.(\ref{mapforvcnlse}). This provide us
more possibility for soliton control, which will be given in a
separate paper.

\mbox{\vspace{1cm}}


{\bf Acknowledgments}

{\noindent \small This work is supported by the NSF of China under
Grant No.10671187 and  10971109. Jingsong He is also supported by
Program for NCET under Grant No.NCET-08-0515. Yishen Li is supported
by NSF of China under Grant No.10971211.  We express our sincere
thanks to Dr. Chaohong Li and Xiwen Guan for their helpful discussion on
early results at Nov. 2007(ANU,Canberra), and Prof. Wuming Liu for
his suggestions at Sept. 2008(USTC,Hefei) and at  Oct.2008(IOP,Beijing).
Jingsong He thanks Prof. Jiefang Zhang(ZJNU,China) for his useful
suggestions at Oct. 2009(NBU,China) on the optical lattice potentials.
Many thanks to Mr. Xiaodong Li, Mr. Jipeng Cheng and Ming Gong(USTC,Hefei) for their
helps. The transformation eq.(\ref{mapforvcnlse})  and its special case
eq.(\ref{mapautonomousnlstonls}) in this paper was presented orally
by Li at ``Integrable System" session, International Conference:
Nonlinear Waves(Beijing, China, June 9-12,2008)(unpublished).
We thank anonymous referee for his/her valuable suggestions and criticisms on the profile of
 single soliton.}




\begin{thebibliography}{200}
\bibitem{cl}H. H. Chen, and C. S. Liu,Solitons in Nonuniform Media, Phys. Rev. Lett.{\bf 37}:693-697(1976).
\bibitem{calgoero1}F. Calogero and  A. Degasperis,Extension of  the
Spectral Transform Method for Solving Nonlinear Evolution
 Equations, Lett. Nuovo Comento {\bf 22}:131-137(1978);
 Extension of  the  Spectral Transform  Method for Solving Nonlinear Evolution
   Equations. II , ibid.{\bf 22}: 263-269(1978).
\bibitem{newell}A. Newell,The general structre of integrable evolution
equations, Proc. R. Soc. Lond. A. {\bf 365}:283-311(1979).
\bibitem{li1}Y.S. Li  and G. C. Zhu,New set of symmetries of the integrable equations,
 Lie algebra and nonisospectral evolution equations. II. AKNS system,J. Phys. A{\bf
 19}: 3713-3725(1986).
\bibitem{akhmediev}N.Akhmediev and  A. Ankiewicz,  {\sl Solitons, Nonlinear
Pulses and Beams},Chapman \& Hall, London,1997; {\sl Dissipative
Solitons},Springer-Verlag, Berlin,2005.
\bibitem{kivshar1}Yu.S. Kivshar and  G. P. Agrawal, {\sl Optical Solitons: From Fibers to Photonic Crystals}(
Academic Press, San Diego, 2003).
\bibitem{vnsah1}V. N. Serkin and  A. Hasegawa,Novel Soliton Solutions of the Nonlinear Schr\"oinger
Equation Model,Phys. Rev. Lett. {\bf 85}:4502-4505(2000).
\bibitem{vnsah2}V. N. Serkin, A. Hasegawa, and T. L. Belyaeva,Nonautonomous Solitons in External Potentials,
Phys. Rev. Lett. {\bf 98}:074102(2007).
\bibitem{kpd}V.I.Kruglov, A.C.Peacock, and J. D. Harvey,Exact Self-Similar Solutions of the Generalized Nonlinear
Schr\"odinger Equation with Distributed Coefficients,
Phys.Rev.Lett.{\bf 90}:113902(2003).
\bibitem{agrawal}S. A. Ponomarenko and G. P. Agrawal,Do Solitonlike Self-Similar Waves Exist in
Nonlinear Optical Media?, Phys. Rev. Lett. {\bf 97}:013901(2006).
\bibitem{kp1}V. V. Konotop and  P.Pacciani,Collapse of Solutions of the Nonlinear Schr\"odinger
Equation with a Time-Dependent Nonlinearity: Application to
Bose-Einstein Condensates, Phys. Rev. Lett. {\bf 94}:240405(2005).
\bibitem{jvv1} J. Belmonte-Beitia, V. M. Perez-Garcia,V. Vekslerchik, and P. J. Torres,
Lie Symmetries and Solitons in Nonlinear Systems with Spatially
Inhomogeneous Nonlinearities, Phys. Rev. Lett. {\bf 98}:064102(2007)
\bibitem{jvv2} J. Belmonte-Beitia, V. M. Perez-Garcia, V. Vekslerchik, and V. V. Konotop,
Localized Nonlinear Waves in Systems with Time- and Space-Modulated Nonlinearities,
 Phys. Rev. Lett. {\bf 100}:164102(2008).
\bibitem{su}H. Saito and M. Ueda, Dynamically Stabilized Bright Solitons in a Two-Dimensional Bose-Einstein
Condensate, Phys. Rev. Lett. {\bf 90}:040403(2003).
\bibitem{liu1}Z. X. Liang, Z.D. Zhang,and  W. M. Liang,Dynamics of a Bright Soliton in Bose-Einstein Condensates
with Time-Dependent Atomic Scattering Length in an Expulsive
Parabolic Potential, Phys. Rev. Lett. {\bf 94}:050402(2005).
\bibitem{sivan}Y. Sivan, G. Fibich, and  M. I. Weinstein,Waves in Nonlinear Lattices: Ultrashort Optical Pulses
and Bose-Einstein Condensates, Phy. Rev. Lett. {\bf 97}:193902(2006)
\bibitem{ss}C.Sulem and  P.L.Sulem, {\sl The Nonlinear Schr\"odinger Equations},Springer-Verlag, New York, 1999.
\bibitem{kontop1}G. Theocharis, Z. Rapti, P. G. Kevrekidis, D.J. Frantzeskakis, and V.V. Konotop,
Modulational instability of Gross-Pitaevskii-type equations in 1+1
dimensions, Phys.Rev. A {\bf 67}:063610(2003).
\bibitem{zhangjiefang1}L.Wu, J.F.Zhang, and L. Li, Modulational instability and bright solitary wave
solution for Bose-Einstein condensates with time-dependent scattering length and harmonic potential,
New J. Phys. {\bf 9}:69(2007).
\bibitem{zhangjiefang2}L.Wu, J.F.Zhang, L.Li,Q.Tian, and K.Porsezian,Similaritons in nonlinear optical systems,
Opt.Express {\bf 16}:6352-6360(2008).
\bibitem{konotop2}V. M.Perez-Garcia, P. J. Torres, and V. V. Konotop,Similarity transformations for nonlinear
Schr\"odinger equations with time-dependent coefficients, Physica D{\bf 221}:31-36(2006).
\bibitem{jbb1}J. Belmonte-Beitia and Gabriel F. Calvo,Exact solutions for the quintic nonlinear Schr\"oinger equation with time and
space modulated nonlinearities and potentials, Phys. Lett. A{\bf 373 }:448-453(2009).
\bibitem{chenlin} S. H. Chen and L. Yi,Chirped self-similar solutions of a generalized nonlinear Schr\"oinger
equation model, Phys. Rev. E{\bf 71}:016606(2005).
\bibitem{he1}J. S. He, M. Ji, and Y.S. Li, Solutions of two kinds of non-isospectral generalized
nonlinear Schr\"odinger equation related to Bose-Einstein
Condensates,Chin. Phys. Lett. {\bf 24}:2157-2160(2007).
\bibitem{ysli}H.M. Li, Y.S. Li, and J. Li,Abundant exact solutions for a strong
dispersion-managed system equation, Chin. Phys. B {\bf 18}:3657-3662(2009).
\bibitem{Gurses}M. G\"urses, Integrable nonautonomous nonlinear Schr\"odinger equations(arXiv:0704.2435v2).
\bibitem{luo1}D.Zhao, H.G. Luo, and H. Y.Chai,Integrability of the Gross-Pitaevskii equation with Feshbach
resonance, Phys. Lett. A{\bf 372}:5644-5650(2008).
\bibitem{luo2}X.G.He, D.Zhao, L.Li, and H.G.Luo,Engineering integrable nonautonomous
nonlinear Schr\"oinger equations, Phys.Rev.E {\bf 79},056610(2009).
\bibitem{kundu} A. Kundu,Integrable nonautonomous nonlinear Schr\"oinger equations are equivalent to
the standard autonomous equation,Phys. Rev. E{\bf 79},015601(R)(2009).
\bibitem{kivshar2}P.J.Louis, E.A. Ostrovskaya, C.M.Savage, and Y.S.Kivshar,
Bose-Einstein condensates in optical lattices:Band-gap structure and solitons, Phys.Rev.A {\bf 67}:013602(2003);
P.J.Louis, E.A. Ostrovskaya, and Y.S.Kivshar,Matter-wave dark solitons in optical, J.Opt.B {\bf 6}:S309-S317(2004);
ibid., Dispersion control for matter waves and gap solitons in optical superlattices, Phys.Rev.A {\bf 71}:
023612(2005).
\bibitem{malomed1} S. K. Adhikari and B. A. Malomed,Tightly bound gap solitons in a Fermi gas,
Eur.Phys. Lett.{\bf 79}:50003(2007); ibid.,Gap solitons in a model of a superfluid fermion gas in optical
lattices,Physica D{\bf 238}:1402-1412(2009);
\bibitem{konotop3}M.Salerno, V.V.konotop, and Yu.V.Bludov, Long-Living Bloch Oscillations of Matter Waves
in Periodic Potentials,Phys. Rev. Lett. {\bf 101}:030405(2008); Yu.V.Bludov, V.V.konotop, and M.Salerno,
Long lived matter waves Bloch oscillations and  dynamical localization by time dependent nonlinearity management,
 J. Phys. B.{\bf 42}:105302(2009) .
\bibitem{kevrekidis1}M. A. Porter, P.G. Kevrekidis, R. Carretero-Gonz$\acute{a}$lez, and D. J. Frantzeskakis,
Dynamics and manipulation of matter-wave solitons in optical superlattices, Phys. Lett.A {\bf 352}:
210-215(2006);R.Carretero-Gonz$\acute{a}$lez, D. J. Frantzeskakis, and P.G. Kevrekidis,Nonlinear waves in
Bose-Einstein condensates: physical relevance and mathematical techniques,Nonlinearity {\bf 21}:R139-R202(2008).
\bibitem{wubiao}Y.P.Zhang and B.Wu, Composition Relation between Gap Solitons and BlochWaves in Nonlinear
Periodic Systems,Phys. Rev. Lett. {\bf 102}:093905(2009).
\bibitem{kivshar3}E.A. Ostrovskaya,Y.S.Kivshar,M. Lisak, B. Hall,and F. Cattani, Coupled-mode theory for
Bose-Einstein condensates'',Phys.Rev.A {\bf 61}:031601(R)(2000).
\bibitem{malomed2}V. S. Shchesnovich, B.A.Malomed, and R.A. Kraenkel, Solitons in Bose-Einstein condensates
trapped in a double-well potential, Physica D{\bf 188}:213-240(2004); T. Mayteevarunyoo, B. A. Malomed,
and G.J.Dong, Spontaneous symmetry breaking in a nonlinear double-well structure, Phys. Rev.A {\bf
78}: 053601(2008).
\bibitem{ketterle1}Y.Shin, M. Saba, A. Schirotzek, T. A. Pasquini, A. E. Leanhardt, D. E. Pritchard and
W. Ketterle,Distillation of Bose-Einstein Condensates in a Double-Well
Potential, Phys. Rev. Lett. {\bf 92}:150401(2004);Y.Shin,G. B. Jo,
M. Saba, T.A. Pasquini, W. Ketterle and D. E. Pritchard,Optical Weak
Link between Two Spatially Separated Bose-Einstein
Condensates,ibdi. {\bf 95}:170402(2005).
\bibitem{carr}D.R.Dounas-Frazer, A.M.Hermundstad,and L.D.Carr,Ultracold Bosons in a Tilted Multilevel
Double-Well Potential,Phys. Rev. Lett. {\bf 99}:200402(2007).
\bibitem{kevrekidis2} G. Theocharis, P.G. Kevrekidis, D. J. Frantzeskakis,and P.Schmelcher, Symmetry breaking
in symmetric and asymmetric double-well potentials,Phys. Rev.
E{\bf 74}: 056608 (2006); C.Wang,P.G. Kevrekidis,N.Whitaker, and
B.A.Malomed, Two-component nonlinear Schr\"odinger models
 with a double-well potential,Physica D{\bf237}:2922-2932(2008);
 C.Wang,P.G. Kevrekidis,N.Whitaker,T.J. Alexander,D. J. Frantzeskakis, and P.Schmelcher,
 Spinor Bose-Einstein condensates in double-well potentials, J. Phys. A{\bf 42}:035201 (2009).
\bibitem{definition}
A partial differential equation(PDE) {\bf I} is said to be
designable integrable if there exist an explicit transformation {\bf
T} with sevearl arbitrary functions  mapping {\bf I} to an
integrable PDE {\bf II}, such that the coefficients of {\bf I} can
be designed artificially and analytically by means of the selection
of arbitrary functions according to different motivations. Here {\bf
II} is a well studied integrable PDE with non-variable coefficents.
In contrast to {\bf I}, {\bf II} is called  rigid integrable.
\bibitem{liu2}X. Q. Liu, S. Jiang, W. B. Fan, and W. M. Liu,Soliton solutions in linear magnetic field and
time-dependent laser field, Commun. Nonlinear Sci. and Numer.
Simul. {\bf 9}:361-365(2004).
\bibitem{xue1}J. K. Xue, Controllable compression of bright soliton matter,J. Phys. B {\bf 38}:
3841-3848(2005).
\bibitem{zhangjiefang3}Q.Yang and J.F.Zhang, Bose-Einstein solitons in time-dependent linear potential,
Optics Communications {\bf 258}:35-42(2006).
\bibitem{kivshar4}D. Poletti,T. J. Alexander,E. A. Ostrovskaya, B.W. Li, and Y. S.  Kivshar,
Dynamics of Matter-Wave Solitons in a Ratchet Potential,Phys. Rev. Lett. {\bf 101}, 150403(2008).
\end{thebibliography}
\end{document}